\documentclass[a4paper,11pt]{article}

\usepackage{jcappub} 
\usepackage[export]{adjustbox} 
\usepackage[utf8]{inputenc}
\usepackage{color}
\usepackage{graphicx}
\usepackage{amsmath,amsfonts,amssymb}
\usepackage{acronym}
\usepackage{xspace}
\usepackage{multirow}
\usepackage{tabularx}
\usepackage{orcidlink}
\usepackage{ragged2e}

\usepackage{setspace}
\usepackage{caption}
\usepackage{subcaption}

\usepackage{comment}

\usepackage{listings}
\usepackage[ruled,vlined]{algorithm2e}

\SetCommentSty{mycommfont}

\newcommand{\wu}{{w_{\rm u}}}

\newcommand{\bea}{\begin{equation}\begin{aligned}} 
\newcommand{\eea}{\end{aligned}\end{equation}}
\newcommand{\be}{\begin{equation}}
\newcommand{\ee}{\end{equation}}

\newcommand{\msun}{M_{\odot}}

\newcommand{\td}{{\rm d}}

\definecolor{rossocorsa}{rgb}{0.83, 0.0, 0.0}
\definecolor{tardisblue}{rgb}{0.0, 0.18, 0.53}

\title{Curbing PBHs with PTAs}

\author[a,b,c]{A.J.~Iovino}
\author[d,e,c]{G. Perna}
\author[c]{A. Riotto}
\author[f]{H.~Veerm\"ae}

\affiliation[a]{Dipartimento di Fisica, ``Sapienza'' Universit\`a di Roma, Piazzale Aldo Moro 5, 00185, Roma, Italy}
\affiliation[b]{Istituto Nazionale di Fisica Nucleare, sezione di Roma, Piazzale Aldo Moro 5, 00185, Roma, Italy}
\affiliation[c]{Department of Theoretical Physics and Gravitational Wave Science Center,  \\
24 quai E. Ansermet, CH-1211 Geneva 4, Switzerland}
\affiliation[d]{Dipartimento di Fisica e Astronomia ``Galileo Galilei'', Universit\`a degli Studi di Padova, Via Marzolo 8, I-35131, Padova, Italy}
\affiliation[e]{INFN, Sezione di Padova, Via Marzolo 8, I-35131, Padova, Italy}
\affiliation[f]{Keemilise ja Bioloogilise F\"u\"usika Instituut, R\"avala pst. 10, 10143 Tallinn, Estonia}

\emailAdd{antoniojunior.iovino@uniroma1.it}
\emailAdd{gabriele.perna@phd.unipd.it}
\emailAdd{antonio.riotto@unige.ch}
\emailAdd{hardi.veermae@cern.ch}

\abstract{Sizeable primordial curvature perturbations needed to seed a population of primordial black holes (PBHs) will be accompanied by a scalar-induced gravitational wave signal that can be detectable by pulsar timing arrays (PTA). We derive conservative bounds on the amplitude of the scalar power spectrum at the PTA frequencies and estimate the implied constraints on the PBH abundance. We show that only a small fraction of dark matter can consist of stellar mass PBHs. The strength and the shape of the constraint depend on the shape of the power spectrum and the nature of the non-Gaussianities. We find that constraints on the PBH abundance arise in the mass range $0.1-10^3\, M_{\odot}$, with the sub-solar mass range being constrained only for narrow curvature power spectra. These constraints are softened when positive non-Gaussianity is introduced and can be eliminated when $f_{\rm NL} \gtrsim 5$.
On the other hand, if the PBH abundance is computed via the theory of peaks, the PTA constraints on PBHs are relaxed, signalling once more the theoretical uncertainties in assessing the PBH abundance. 
We further discuss how strong positive non-Gaussianites can allow for heavy PBHs to potentially seed supermassive BHs.}

\begin{document}
\maketitle
\flushbottom
\section{Introduction}

Primordial black holes (PBHs)~\cite{Zeldovich:1967lct, Hawking:1971ei, Carr:1974nx} are currently in the spotlight as they may solve several open questions in astrophysics and cosmology. Different mechanisms and models have been proposed as viable for PBH production and,
depending on the formation scenario considered and on the models for PBH formations, a variety of mass functions have been predicted (see Ref.~\cite{LISACosmologyWorkingGroup:2023njw} for a recent review). Throughout this work, we will assume the standard formation scenario which suggests that PBHs form out of the 
gravitational collapse after the horizon re-entry of large over-densities in the primordial density contrast field~\cite{Ivanov:1994pa,Ivanov:1997ia,Blinnikov:2016bxu}. In such a scenario, a non-negligible PBH abundance requires an enhanced primordial curvature power spectrum at scales smaller than the scale of the Cosmic Microwave Background (CMB), at which the spectral amplitude is around $10^{-9}$~\cite{Planck:2018jri}. 

In addition to producing PBHs, the same primordial curvature perturbations generate tensor perturbations, also known as scalar-induced Gravitational Waves (SIGW). As first shown by Refs.~\cite{Matarrese:1993zf,Acquaviva:2002ud,Mollerach:2003nq,Ananda:2006af,Baumann:2007zm}, when going to second order, scalar and tensor perturbations are coupled and the former provide a source for Gravitational Waves (GW) (see also Refs.~\cite{Espinosa:2018eve,Domenech:2021ztg}). The statistical properties of the curvature fluctuations can leave an imprint on this signal. Indeed, the presence of primordial non-Gaussianity (NG)~\cite{Gangui:1993tt,Bartolo:2001cw,Maldacena:2002vr,Bartolo:2003jx,Bartolo:2004if,Celoria:2018euj} can modify the spectrum inducing specific features depending on the amplitude of the sourcing scalar power spectrum~\cite{Garcia-Saenz:2022tzu, Unal:2018yaa, Adshead:2021hnm,Abe:2022xur, Perna:2024ehx}. Moreover, the specific features of the spectra and the NG modifications, as well as the amplitude of the induced SIGW background, depend not only on the shape of the curvature power spectrum considered, but also on the specific scales involved (see, e.g.,~\cite{Cai:2018dig, Cai:2019elf}). Hence detectable SIGWs offer a potential way to probe early universe cosmology.

The recent Pulsar Timing Array (PTA)  data release by the NANOGrav~\cite{NANOGrav:2023gor, NANOGrav:2023hde}, EPTA (in combination with InPTA)~\cite{EPTA:2023fyk, EPTA:2023sfo, EPTA:2023xxk}, PPTA~\cite{Reardon:2023gzh, Zic:2023gta, Reardon:2023zen} and CPTA~\cite{Xu:2023wog} collaborations shows evidence of a Hellings-Downs pattern in the angular correlations which is characteristic of GWs. The most stringent constraints and largest statistical evidence come from the NANOGrav 15-year data (NANOGrav15). Even though it is currently not possible to pinpoint the sources of this signal, this discovery is of extreme importance since it opens a further window into the early universe and has an impact on the GW search at the nHz scale (see, e.g.,~\cite{Ellis:2023oxs, Figueroa:2023zhu}).

Since PTAs are sensitive to frequencies of the order of the nHz, they are associated with the production of PBHs in the stellar mass range~\cite{Chen:2019xse,DeLuca:2020agl,Vaskonen:2020lbd,Kohri:2020qqd,Domenech:2020ers,Dandoy:2023jot,Franciolini:2023pbf,Wang:2023ost}. Crucially, requiring that the produced SIGW background does not exceed the one registered by the NANOGrav15, strongly limits the maximum amplitude of the curvature power spectrum, and thereby the abundance of PBHs in the corresponding mass range. 
Additionally, if the power spectrum is enhanced at scales larger than accessible by PTAs, PBHs can serve as seeds of supermassive black holes (SMBHs)~\cite{Duechting:2004dk, Bernal:2017nec, Serpico:2020ehh}. Nevertheless, in this case, the scales related to the formation process are strongly constrained by CMB spectral distortion analysis of the FIRAS collaboration~\cite{Fixsen:1996nj,Chluba:2012gq,Chluba:2012we,Chluba:2013dna,Bianchini:2022dqh} and, without NGs, the allowed PBH abundance is too small to provide a primordial origin for SMBHs seeds.

In this work constraints on PBHs arising from NANOGrav15 are considered in detail with an emphasis on local primordial NGs and assuming different shapes for the primordial power spectrum. After reviewing the origin of NGs in the standard formation scenario for PBHs in Sec.~\ref{sec:NGs}, in Sec.~\ref{sec:Abu} we discuss different formalisms for estimating the PBH abundance and the role of NGs. In Sec.~\ref{sec:SIGWs} we summarise the computation of the SIGW background. Then, in Sec.~\ref{sec:Anal}, we derive PTA constraints on the power spectrum and the abundance of PBHs, and comment on how large positive NGs affect the generation of PBHs as seeds for SMBHs. We conclude in Sec.~\ref{sec:Conc}.

\section{Non-Gaussian primordial curvature perturbations}
\label{sec:NGs}

Generally, different models for PBH production require an enhancement of the curvature power spectrum that depends on the model considered. Typical shapes widely used in literature are the \emph{log-normal} (LN)
\be\label{eq:PSlog}
    \mathcal{P}_\zeta(k)
    =\frac{{A}}{\sqrt{2 \pi} \Delta} \exp \left(-\frac{1}{2 \Delta^2} \ln ^2\left(\frac{k}{ k_*}\right)\right)\,,
\ee
where $k_*$ denotes the peak frequency, or the \emph{broken power-law} (BPL) 
\be \label{eq:PPL}
    \mathcal{P}_{\zeta}(k)
    = {A} \,\frac{\left(\alpha+\beta\right)^{\lambda}}{\left[\beta\left(k / k_*\right)^{-\alpha/\lambda}+\alpha\left(k / k_*\right)^{\beta/\lambda}\right]^{\lambda}},
\ee
where the parameters $\alpha, \beta>0$ describe the growth and decay of the spectrum around the peak and $\lambda$ the width of the peak. Typically, $\alpha \approx 4$~\cite{Byrnes:2018txb,Karam:2022nym} and $\beta \lesssim 4$. The cases considered in this work are reported in Tab.~\ref{tab:1}.
\begin{table}[h!]
    \centering
    \begin{tabular}{p{1.2cm}p{5cm}p{1.5cm}}
         \hline \hline
         Case &  PS shape &   $f_{\rm NL}$  \\
         \hline\hline
         LN & LN $(\Delta=0.5)$ & $[-2,10]$ \\\\
         
         BPL1& BPL $(\alpha=4,\beta=3,\lambda=1)$&  $[-2,10]$ \\\\
         
         BPL2& BPL $(\alpha=4,\beta=0.5,\lambda=1)$&  $[-2,10]$ \\
         \hline \hline
    \end{tabular}
    \caption{\justifying List of benchmark cases considered in this work.}
    \label{tab:1}
\end{table}

The standard formation scenario assumes that PBHs arise from the gravitational collapse of significant over-densities in the primordial density contrast field~\cite{Ivanov:1994pa,Ivanov:1997ia}. Given the random nature of this field, the calculation of PBH abundance is inherently statistical and requires precise knowledge of the probability density function (PDF) of the density fluctuations.

Limiting the analysis to the assumption that the PDF of density fluctuations $\delta$ follows the Gaussian statistics is theoretically flawed for two reasons.  
First, density fluctuations in the primordial radiation field originate from curvature perturbations $\zeta$, previously stretched on super-horizon scales during inflation, after their horizon re-entry. 
In the long-wavelength approximation, the equation which relates curvature perturbations to density fluctuations is intrinsically non-linear~\cite{Harada:2015yda}
\begin{align}\label{eq:NonLinearDelta}
    \delta(\vec{x},t) \!=\!  
    - \frac23\frac{\Phi}{(aH)^2}
    e^{-2 \zeta(\vec{x})}
    \bigg[
    \nabla^2\zeta(\vec{x}) \!+\! \frac{1}{2} \partial_{i}\zeta(\vec{x})
 \partial_{i}\zeta(\vec{x})
    \bigg],\!\!
\end{align}
where $a$ denotes the scale factor, $H$ the Hubble rate, $\Phi$ is related to the equation of state parameter $\wu$ of the universe. For $\wu$ constant, $\Phi= 3(1+\wu)/(5+3\wu)$~\cite{Polnarev:2006aa}. We dropped the explicit $\vec{x}$ and $t$ dependence for brevity.
For this reason, even under the assumption that curvature perturbations follow exact Gaussian statistics, density fluctuations inherit an unavoidable amount of NG from non-linear (NL) corrections~\cite{DeLuca:2019qsy,Young:2019yug,Germani:2019zez}.
Second, there is no guarantee that $\zeta$ is a Gaussian field -- we refer to such cases as \textit{primordial} NGs. Local primordial NGs can be described by a relation  
\be\label{eq:zeta}
    \zeta(\vec{x}) = F(\zeta_{\rm G}(\vec{x}))
\ee
between $\zeta$ and its Gaussian counterpart $\zeta_G$, which depends on the mechanism that generates the enhanced power spectrum at small scales (see the related discussion in Ref.~\cite{Ferrante:2022mui}). Such NGs are thus generically independent of large-scale NGs constrained by CMB data (e.g.~\cite{Planck:2019kim}).
Copious PBH production via critical collapse of large overdensities is achievable in a wide range of scenarios including single-field inflation with
an inflation point in the inflaton’s potential (usually called USR models)~\cite{ Garcia-Bellido:2017mdw,Pi:2017gih, Kannike:2017bxn,Ballesteros:2020qam,Inomata:2016rbd,Iacconi:2021ltm,Kawai:2021edk,Bhaumik:2019tvl,Cheong:2019vzl,Inomata:2018cht,Dalianis:2018frf,Motohashi:2019rhu,Hertzberg:2017dkh,Ballesteros:2017fsr,Karam:2022nym,Rasanen:2018fom,Balaji:2022rsy,Frolovsky:2023hqd,Dimopoulos:2017ged,Germani:2017bcs,Choudhury:2013woa,Ragavendra:2023ret,Cheng:2021lif,Franciolini:2023lgy,Karam:2023haj,Mishra:2023lhe,Cole:2023wyx,Karam:2023haj,Frosina:2023nxu,Franciolini:2022pav,Choudhury:2024one,Wang:2024vfv,Stamou:2021qdk,Stamou:2024lqf,Heydari:2021gea,Heydari:2021qsr,Heydari:2023xts,Pi:2022ysn} and models with spectator field, i.e., the curvaton~\cite{Enqvist:2001zp,Lyth:2001nq,Sloth:2002xn,Lyth:2002my,Dimopoulos:2003ii,Kohri:2012yw,Kawasaki:2012wr,Kawasaki:2013xsa,Carr:2017edp,Ando:2017veq,Ando:2018nge,Chen:2019zza,Liu:2020zzv,Pi:2021dft,Cai:2021wzd,Liu:2021rgq,Chen:2023lou,Torrado:2017qtr,Chen:2023lou,Cable:2023lca,Inomata:2023drn}.

To be as model-independent as possible we follow the common approach and describe $\zeta$ by the usual power series expansion
\be\label{eq:exp}
    \zeta 
    = \zeta_{\rm G} 
    + \frac{3}{5}f_{\rm NL}(\zeta_{\rm G}^2 -\langle \zeta_{\rm G}^2 \rangle)
    + \frac{9}{25}g_{\rm NL}\zeta_{\rm G}^3 + \dots\,,
\ee
where $\zeta_{\rm G}$ obeys the Gaussian statistics while the coefficients $f_{\rm NL}$, $g_{\rm NL}$ encode deviations from the Gaussian limit. In general, the coefficients can depend on
the scale of the perturbation. We omit this possibility here.

\section{PBH formation}\label{sec:Abu}

\begin{figure}[t]
    \centering
    \includegraphics[width=0.95\textwidth]{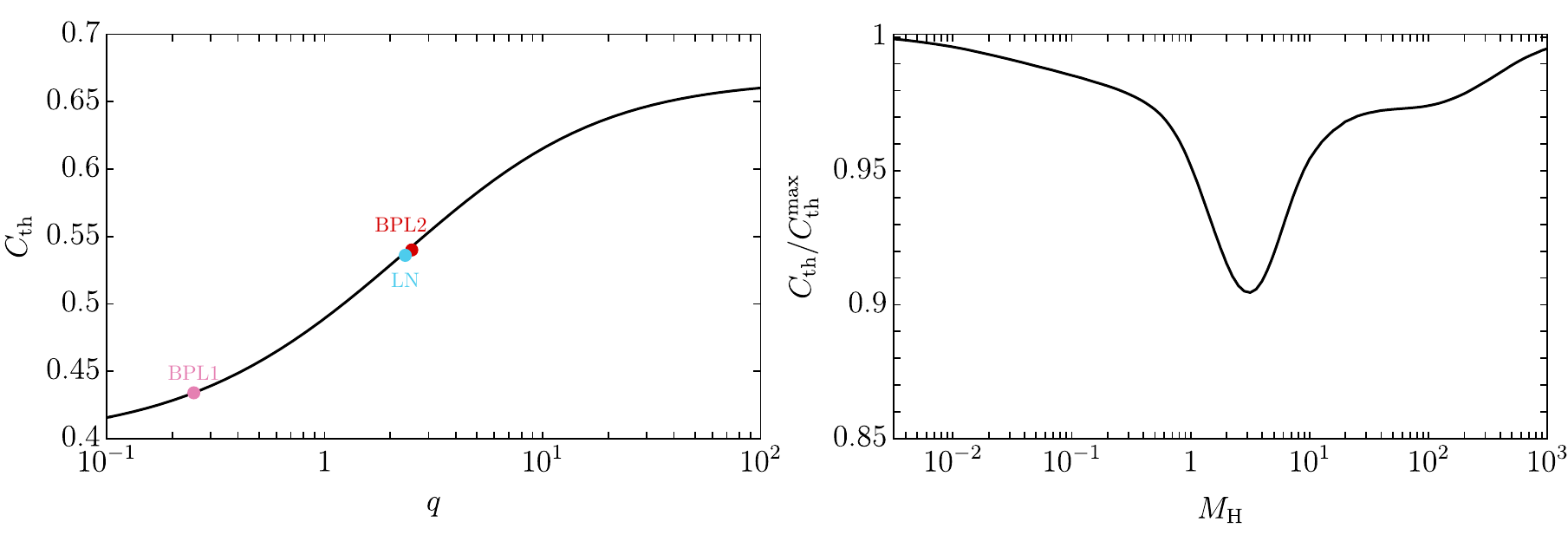}
    \caption{\justifying 
\textit{Left panel:} Threshold $C_{\rm th}$ as a function of the shape parameter $q$. The coloured dots indicate the three benchmark scenarios in Table~\ref{tab:1}. \textit{Right panel:} Relative change of threshold $C_{\rm th}$ as a function of the horizon mass $M_{\rm H}$.
}
\label{fig:thre}
\end{figure}

A PBH is formed when a sufficiently large primordial overdensity collapses after horizon re-entry. The mass of the resulting PBH follows a critical scaling law
\be
    M_{\rm PBH} = \mathcal{K} M_k (\mathcal{C} - \mathcal{C}_{\rm th})^{\gamma}\,,
\ee
where
\be \label{eq:M_k}
    M_k
    \approx 17 \msun \left(\frac{k}{10^6{\rm Mpc}^{-1}} \right)^{-2}\, \left(\frac{g_{*}}{10.75}\right)^{-1/6} \,
\ee  
is the mass contained within a Hubble horizon corresponding to a comoving scale $k$. The threshold $\mathcal{C}_{\rm th}$ is estimated following Ref.~\cite{Musco:2020jjb} and depends on the shape of the power spectrum $q$. In Fig.\,\ref{fig:thre} we show the variation of the threshold as a function of the horizon mass due to the QCD phase transition and its dependence on the spectral shape parameter $q$.
We also account for the time dependence of $\mathcal{C}_{\rm th}$, the critical exponent $\gamma$ and the prefactor $\mathcal{K}$ during the QCD phase transition~\cite{Musco:2023dak}. We fix $\mathcal{K}=4.4$ and $\gamma=0.38$.

The fraction of energy density $\beta_k(M_{\rm PBH})\td \ln M_{\rm PBH}$ collapsing into PBHs can be estimated as 
\be\label{eq:betak}
    \beta_k(M_{\rm PBH})
    = \int_{\mathcal{C}_{\rm th}} \! \td\mathcal{C} \, P_k(\mathcal{C}) \frac{M_{\rm PBH}}{M_k}  \delta\left[ \ln\frac{M_{\rm PBH}}{M_{\rm PBH}(\mathcal{C})} \right]\!,
\ee
where $P_k(\mathcal{C})$ denotes the probability a BH will form in the Hubble patch. The PBH mass function can be obtained directly from the collapse fraction:
\bea\label{eq:df_PBH}
    \!\frac{\td f_{\rm PBH}}{\td \ln M_{\rm PBH}}
&   \!=\! \frac{1}{\Omega_{\rm DM}}\int \frac{\td M_k}{M_k} \, \beta_k(M_{\rm PBH} ) \left(\frac{M_{\rm eq}}{M_k}\right)^{1/2} \!\!,
\eea
where $M_{\rm eq} \approx 2.8\times 10^{17}\,\,M_{\odot}$ is the horizon mass at the time of matter-radiation equality and $\Omega_{\rm  DM} = 0.12h^{-2}$ is the cold dark matter density~\cite{Planck:2018jri}. To characterise the PBH population, we will consider the PBH abundance and the mean PBH mass,
\bea
   f_{\rm PBH} 
   &= \int \frac{\td M_{\rm PBH}}{M_{\rm PBH}} \frac{\td f_{\rm PBH}}{\td \ln M_{\rm PBH}}\, , \\
   \langle M_{\rm PBH} \rangle 
   &= f_{\rm PBH} \left(\int \frac{\td M_{\rm PBH}}{M_{\rm PBH}^2} \frac{\td f_{\rm PBH}}{\td \ln M_{\rm PBH}}\right)^{-1} \,,
\eea
where the mean PBH mass is computed with respect to the PBH number density (see e.g.~\cite{Andres-Carcasona:2024wqk}). This offers a clear basis for comparing constraints on PBHs with extended mass functions.

The above description can be applied to estimate PBH abundance using both the threshold statistics and peaks theory. However, the shape and the origin of $P_k(\mathcal{C})$ in these two approaches are different.

\paragraph{Threshold statistics} 
dictates that the PBH formation probability can be estimated from the statistics of the compaction function $\mathcal{C}$~\cite{Ferrante:2022mui,Gow:2022jfb}, generically defined as twice the local mass excess over the areal radius\footnote{Using the threshold statistics approach, which relies on average compaction profiles~\cite{Ianniccari:2024bkh}, corrections to the horizon crossing and from the non-linear radiation transfer function~\cite{Franciolini:2023wun,DeLuca:2023tun} can affect the PBH abundance. Since the amount of these corrections is still not well understood we leave their inclusions for future work.}. The latter can be estimated in scenarios with NG curvature perturbations $\zeta$ in case they can be characterized via an auxiliary Gaussian field $\zeta_G$ as outlined in Sec.~\ref{sec:NGs}. The compaction function $\mathcal{C} = \mathcal{C}_1 - \mathcal{C}_1^2/(4\Phi)$ can then be constructed from
$\mathcal{C}_1 = \mathcal{C}_{\rm G} \, \td F/\td\zeta_{\rm G}$ which can be expressed in terms of the Gaussian field $\mathcal{C}_{\rm G} = -2\Phi\,r\,\zeta_{\rm G}^{\prime}$.

The PBH formation probability in a Hubble patch is then
\be\label{eq:pk}
    P_k(\mathcal{C}) = 
    \int_{\mathcal{D}} \delta(\mathcal{C} - \mathcal{C}(\mathcal{C}_{\rm G},\zeta_{\rm G}))
    P_{{\rm G},k}(\mathcal{C}_{\rm G},\zeta_{\rm G})\td\mathcal{C}_{\rm G} \td\zeta_{\rm G}\,,
\ee
where the domain of integration
$\mathcal{D} =
\left\{
    \mathcal{C}(\mathcal{C}_{\rm G},\zeta_{\rm G}) > \mathcal{C}_{\rm th}  
    ~\land~\mathcal{C}_1(\mathcal{C}_{\rm G},\zeta_{\rm G}) < 2\Phi
\right\}$.
and the Gaussian components are distributed as
\be
    P_{\mathrm{G}}\left(\mathcal{C}_{\mathrm{G}}, \zeta_{\mathrm{G}}\right)
    = \frac{e^{\left[-\frac{1}{2\left(1-\gamma_{c r}^2\right)}\left(\frac{\mathcal{C}_{\mathrm{G}}}{\sigma_c}-\frac{\gamma_{c r} \zeta_{\mathrm{G}}}{\sigma_r}\right)^2 \!-\! \frac{\zeta_{\mathrm{G}}^2}{2 \sigma_r^2}\right]}}{2 \pi \sigma_c \sigma_\tau \sqrt{1-\gamma_{c r}^2}}.
\ee
The correlators are given by
\begin{subequations}
\begin{align}
    & \sigma_c^2=\frac{4 \Phi^2}{9} \int_0^{\infty} \frac{\td k}{k}\left(k r_m\right)^4 W^2\left(k, r_m\right)P^{T}_\zeta
    \,, \\
    &\sigma_{c r}^2=\frac{2 \Phi}{3} \! \int_0^{\infty} \!\! \frac{\td k}{k} \!\left(k r_m\right)^2 \!W\!\!\left(k, r_m\right) \!W_s\!\left(k, r_m\right) \!P^{T}_\zeta\! 
    \,, \\
    &\sigma_r^2=\int_0^{\infty} \frac{\td k}{k} W_s^2\left(k, r_m\right) P^{T}_\zeta \,,
\end{align}
\end{subequations}
with $P^{T}_\zeta=T^2\left(k, r_m\right) P_\zeta(k)$, and  $\gamma_{c r} \equiv \sigma_{c r}^2 / \sigma_c \sigma_\tau$.
We have defined $W\left(k, r_m\right), $ $W_s\left(k, r_m\right)$ and $T\left(k, r_m\right)$ 
as the top-hat window function, the spherical-shell window function, and the radiation transfer function~\cite{Young:2022phe}\footnote{The choice of the window function is dictated by the approach used to evaluate the threshold for the critical collapse. Using different window functions is possible, but it would be necessary to reevaluate the threshold in order to maintain the accuracy of the evaluation~\cite{Young:2019osy}.}.

The PBH mass function \eqref{eq:df_PBH} can be expressed as 
\begin{align}\label{eq:df_PBH_th}
    &\frac{\td f_{\rm PBH}}{\td \ln M_{\rm PBH}} 
    \!=\! \frac{1}{\Omega_{\rm DM}} \int_{M_{\rm H}^{\rm min}}
    \frac{d M_{\rm H}}{M_{\rm H}} \left(\frac{M_{\rm eq}}{M_{\rm H}}\right)^{1/2}\left(\frac{M_{\rm PBH}}{\mathcal{K} M_{\rm H}}\right)^\frac{1+\gamma}{\gamma} \nonumber \\
    &\times \frac{{\cal K}}{\gamma\sqrt{\Lambda}} \int \td\zeta_{{\rm G},k} P_{\rm G}({\cal C}_{\rm G}(M_{\rm PBH},\zeta_{\rm G}), \zeta_{\rm G}) \left(\frac{dF}{d\zeta_{\rm G}}\right)^{-1}\!\!\!\!,
\end{align}
where $\Lambda=1 - ({\cal C}_{\rm th} - \left(M_{\rm PBH}/(\mathcal{K}M_H)\right)^{1/\gamma})/\Phi$.

\paragraph{Theory of peaks} 
It is well known that the Press-Schechter approach does not agree with the theory of peaks (see, e.g., Refs.~\cite{Green:2004wb, Young:2014ana, DeLuca:2019qsy}).
A technical drawback of peaks theory is that it is not clear how to include NGs in the computation of the abundance in this approach using a generic functional form for the curvature perturbation field as in Eq.~\eqref{eq:zeta}. Hence, for comparison, we limit our analysis for the peak theory approach to the case of negligible primordial NGs\footnote{The impact of local-type NG on PBH abundance in peak theory is expected to be reduced,  especially for higher-order terms \cite{Young:2022phe} and for peaked power spectra. Indeed,  the compaction is volume-averaged over the scale of the perturbation and depends on the curvature perturbation at its edge, rather than its centre.}.

In this approach, the starting point is the number density of peaks with a height in the interval $(\nu,\nu+\td \nu)$. When $\nu \gg 1$, it is given by~\cite{Bardeen:1985tr}
\be
    \frac{\td \mathcal{N}}{\td \nu}
    =\frac{1}{4 \pi^2 r_m^3}\left(\frac{\sigma_{cc}}{\sigma_{c}}\right)^3 \nu^3 \exp \left(-\frac{\nu^2}{2}\right)\,,
\ee
where we introduced the rescaled peak height $\nu \equiv \mathcal{C}_1 / \sigma_c$ and the first rescaled moment of the distribution~\cite{Balaji:2023ehk}
\be
    \sigma_{cc}^2 = \frac{4 \Phi^2}{9} \int_0^{\infty} \frac{\td k}{k}\left(k r_m\right)^6 W^2\left(k, r_m\right)P^{T}_\zeta\,.
\ee
The number of peaks within a Hubble volume is therefore
\be
    \frac{\td N_k}{\td \nu} 
    = \frac{4\pi}{3} r_m^3 \mathcal{N}
\ee
and, since $N_k \ll 1$, the probability of finding at least one peak within a range $(\nu,\nu+\td \nu)$ within a Hubble volume is $P_k \approx \td N_k$.
The mass fraction of PBHs is then~\cite{Yoo:2018kvb,Yoo:2019pma,Gow:2020bzo,Franciolini:2022tfm}
\be
    P_k(\mathcal{C}) 
    = \frac{1}{3 \pi}\left(\frac{\sigma_{cc}}{\sigma_{c}}\right)^3 \nu^3 \exp \left(-\frac{\nu^2}{2}\right)
\ee
and the domain of integration agrees with Eq.~\ref{eq:betak}. Thus, peaks theory predicts an additional $\nu^3$ factor to the collapse probability in the Gaussian case.

Recasting the quantities in terms of $M_k$, as in~\eqref{eq:df_PBH_th}, we can express the PBH abundance as
\begin{align}\label{eq:df_PBH_peak}
    \frac{\td f_{\rm PBH}}{\td \ln M_{\rm PBH}} 
    = &\frac{1}{\Omega_{\rm DM}} \int_{M_{\rm H}^{\rm min}}
    \frac{d M_{\rm H}}{M_{\rm H}} \left(\frac{M_{\rm eq}}{M_{\rm H}}\right)^{1/2}\left(\frac{M_{\rm PBH}}{\mathcal{K} M_{\rm H}}\right)^\frac{1+\gamma}{\gamma} 
    \nonumber\\
    & \times \frac{{\cal K}}{\gamma} \frac{\left(2 \Phi \left(1-\sqrt{\Lambda}\right)\right)^3}{3\pi \sigma_c^4\Lambda^{1/2}} \left(\frac{\sigma_{cc}}{\sigma_c}\right)^3 
     \exp\left[{\frac{-2\Phi^2}{\sigma_c^2}\left(1-\sqrt{\Lambda}\right)^2}\right]\,.
\end{align}

\section{Scalar-induced gravitational waves}
\label{sec:SIGWs}

Primordial curvature fluctuations large enough to generate PBHs will also induce a non-negligible SIGW background. At the second order in perturbation theory, scalar modes can source tensor modes (for a review, see Ref.~\cite{Domenech:2021ztg}). Starting from Einstein's equations and omitting anisotropic stress, the equations of motion for GWs in Fourier space are
\be\label{eom1}
    h''_{\lambda}(\textbf{k}, \eta) + 2\mathcal{H}h'_{\lambda}(\textbf{k}, \eta) + k^2h_{\lambda}(\textbf{k}, \eta) = 4\mathcal{S}_{\lambda}(\textbf{k}, \eta)\,, 
\ee
with $\lambda$ indicating the polarization and the $'$ being the derivative with respect to the conformal time $\eta$. The source term on the right-hand side reads~\cite{Baumann:2007zm}
\be\label{ftsource}
    \mathcal{S}_{\lambda}(\textbf{k}, \eta)
    \!=\! \int\frac{\td^3\textbf{q}}{(2\pi)^{\frac32}}Q_{\lambda}(\textbf{k},\textbf{q})f(|\textbf{k}-\textbf{q}|,q,\eta)\zeta_\textbf{q}\zeta_{\textbf{k}-\textbf{q}}\,,
\ee
where $Q$ is a projection factor and $f$ is a function that contains the linear evolution of the Newtonian potential after horizon re-entry and the transfer function mapping them to the curvature perturbation $\zeta$. Eq.~\eqref{eom1} can be solved using the Green's functions approach
\be
    h_{\lambda}(\textbf{k}) \!=\! 4\int\frac{\td^3\textbf{q}}{(2\pi)^{\frac32}} Q_{\lambda}(\textbf{k},\textbf{q})I(|\textbf{k}-\textbf{q}|,q,\eta)\zeta_\textbf{q}\zeta_{\textbf{k}-\textbf{q}} \,,
\ee
where $I$ is the transfer function. To speed up our analysis, we assume perfect radiation domination and do not account for the variation of sound speed during the QCD era (see, for example,~\cite{Hajkarim:2019nbx, Abe:2020sqb}), which also leads to specific imprints in the low-frequency tail of any cosmological SGWB~\cite{Franciolini:2023wjm}. 
On top of that, cosmic expansion may additionally be affected by unknown physics in the dark sector, which can, e.g., lead to a brief period of matter domination of kination~\cite{Ferreira:1997hj,Pallis:2005bb,Redmond:2018xty,Cai:2020qpu,Co:2021lkc,Gouttenoire:2021jhk,Chang:2021afa,Cai:2023uhc,Domenech:2024rks}. Explicit analytical expressions for the kernel in different epochs can be found in~\cite{Domenech:2019quo}, after performing the oscillation average well inside the horizon.

The dimensionless GW power spectrum $\mathcal{P}_h(k)$ is defined as
\be
    \langle h_{\lambda_{1}}(\mathbf{k}_1)h_{\lambda_{2}}(\mathbf{k}_2) \rangle = \delta^3(\mathbf{k}_1 + \mathbf{k}_2) \delta_{\lambda_1\lambda_2} \frac{2\pi^2}{k^3} \mathcal{P}_{h,\lambda_1}(k_1)
\ee
and hence 
\begin{align}
    \Omega_{\rm GW}(k,\eta) 
    = \frac{1}{48} \left(\frac{k}{a(\eta)H(\eta)}\right) \sum_{\lambda = +,\times} \overline{\mathcal{P}_{h,\lambda}(k)}
\end{align}
at the production. In the Gaussian case, one obtains
\begin{align}
    \Omega_{\rm GW}(k,\eta)|_{\rm G} =& \hspace{0.1cm}\frac{1}{12}  \left(\frac{k}{a(\eta)H(\eta)}\right)^2 \int_{0}^{\infty} {\rm{d}}t \int_{-1}^{1} {\rm{d}}s \hspace{0.1cm}\frac{1}{u^2 v^2}\overline{\Tilde{J}^2(u,v,x)}\hspace{0.1cm}
\mathcal{P}_{\zeta_g}(v k)\mathcal{P}_{\zeta_g}(u k)\,,
    \label{Eq:Gaussian}
\end{align}
where $x = k\eta$, $t$ and $s$ are dimensionless variables introduced to ease the numerical analysis, $\tilde{J}$ is related to the transfer function $I$ defined before and the overbar indicates an oscillation average after the modes are well inside the horizon. Furthermore $u=(t+s+1)/2$ and $v=(t-s+1)/2$. The spectrum observed today can be obtained using entropy conservation
\be
    h^2\Omega_{\rm GW}|_{k,0} = h^2\Omega_{\rm rad,0} \left(\frac{g_{s,\eta_0}}{g_{s,\eta_k}}\right)^{\frac{4}{3}} \frac{g_{\rho,\eta_k}}{g_{\rho,\eta_0}} \Omega_{\rm GW} (k,\eta)\,,
\ee
where we used the energy ($g_{\rho}$) and entropy density ($g_s$) degrees of freedom of Ref.~\cite{Borsanyi:2016ksw}. According to Ref.~\cite{Planck:2018vyg}, we consider $h^2\Omega_{\rm rad,0}=4.2 \cdot 10^{-5}$.

\subsection{Effects of primordial NG on SIGW}

As shown in~\cite{Kohri:2018awv, Cai:2018dig, Cai:2019elf, Hajkarim:2019nbx, Atal:2021jyo, Yuan:2020iwf, Domenech:2021and, Garcia-Saenz:2022tzu, Liu:2023ymk,  Yuan:2023ofl, Li:2023xtl, Unal:2018yaa, Adshead:2021hnm,Abe:2022xur, Perna:2024ehx} primordial NGs can leave a non-negligible imprint on the SIGW power spectrum. Hence, considering only the quadratic contribution $f_{\rm NL}$ in the local expansion in Eq.~\eqref{eq:exp}\footnote{We specify that, in principle, all the terms in the power series expansion in Eq.~\eqref{eq:exp} should be included. Being interested mainly in the qualitative effect of NGs, we truncated the power series at the quadratic contribution, $f_{\rm NL}$, setting all other terms to 0. These additional NG corrections would further shift the results in Fig.~\ref{fig:GWfPBHNGs}.}, at the next-to-leading order one obtains (see, e.g.,~\cite{Unal:2018yaa,Adshead:2021hnm})
\begin{align}
    \label{eq:t}
    \Omega_{\rm GW}(k,\eta)|_{\rm t} 
    &= \frac{1}{12\pi} \left(\frac{k}{a(\eta)H(\eta)}\right)^2 f_{\rm NL}^2
    \int_{0}^{\infty} {\rm{d}} t_1 \int_{-1}^{1} {\rm{d}}s_1 \int_{0}^{\infty} {\rm{d}} t_2 \int_{-1}^{1} {\rm{d}}s_2\int_{0}^{2\pi}{\rm{d}}\varphi_{12}
     \\
    \times&\cos 2\varphi_{12}\frac{u_1v_1}{(u_2v_2)^2}\frac{1}{w_{a,12}^3}\overline{\Tilde{J}(u_1,v_1,x)\Tilde{J}(u_2,v_2,x)}
    \mathcal{P}_{\zeta_g}(v_2k)\mathcal{P}_{\zeta_g}(u_2k)\mathcal{P}_{\zeta_g}(w_{a,12} k)\nonumber\,,
    \\
    \label{eq:u}
    \Omega_{\rm GW}(k,\eta)|_{\rm u} &= \hspace{0.1cm}\frac{1}{12\pi} \left(\frac{k}{a(\eta)H(\eta)}\right)^2 f_{\rm NL}^2
    \int_{0}^{\infty} {\rm{d}} t_1 \int_{-1}^{1} {\rm{d}}s_1 \int_{0}^{\infty} {\rm{d}} t_2 \int_{-1}^{1} {\rm{d}}s_2\int_{0}^{2\pi}{\rm{d}}\varphi_{12} 
    \\
    \times& \cos 2\varphi_{12} \frac{u_1u_2}{(v_1v_2)^2}\frac{1}{w_{b,12}^3}\overline{\Tilde{J}(u_1,v_1,x)\Tilde{J}(u_2,v_2,x)}
    \mathcal{P}_{\zeta_g}(v_1k)\mathcal{P}_{\zeta_g}(v_2k)\mathcal{P}_{\zeta_g}(w_{b,12} k)\nonumber
    \\
    \label{eq:hyb}
    \Omega_{\rm GW}(k,\eta)|_{\rm hyb} 
    &=\hspace{0.1cm} \frac{1}{12} \left(\frac{k}{a(\eta)H(\eta)}\right)^2 f_{\rm NL}^2\int_0^{\infty}{\rm{d}}t_1\int_{-1}^1 {\rm{d}}s_1\int_0^{\infty} {\rm{d}}t_2\int_{-1}^1 {\rm{d}}s_2
    \\
    \times& \frac{1}{(u_1 v_1 u_2 v_2)^2} \hspace{0.1cm} \overline{\Tilde{J}^2(u_1,v_1,x)}
    \mathcal{P}_{\zeta_g}(u_1 k)\mathcal{P}_{\zeta_g}(v_2v_1 k)\mathcal{P}_{\zeta_g}(u_2v_1 k)\nonumber\,.
\end{align}
As in the Gaussian case, the $u_i$ and $v_i$ are related similarly to the integration variables $t_i$ and $s_i$. 
The GW spectrum can then be expressed as
\bea\label{eq:total_SIGW}
    \Omega_{{\rm GW}}|_{k,0} 
    &= A^2 \Omega_{\rm GW}|_{{\rm G}}(k/k_*,0) + A^3 f_{\rm NL}^2\Omega_{\rm GW}|_{{\rm NG}}(k/k_*,0)\,,
\eea
where $\Omega_{\rm GW}|_{{\rm NG}}$ is the sum of \eqref{eq:t}, \eqref{eq:u} and \eqref{eq:hyb} with $A=1$, $f_{\rm NL} = 1$ computed assuming $k_* = 1$.\footnote{Factoring out $k_*$ dependence in this way is possible in pure radiation dominance, that is, when QCD effects are neglected.} In this way, given the spectral shape is fixed, the GW spectrum can be straightforwardly computed for given $A$, $k_*$ and $f_{\rm NL}$.

\begin{figure*}
    \centering
    \includegraphics[width=0.99\textwidth]{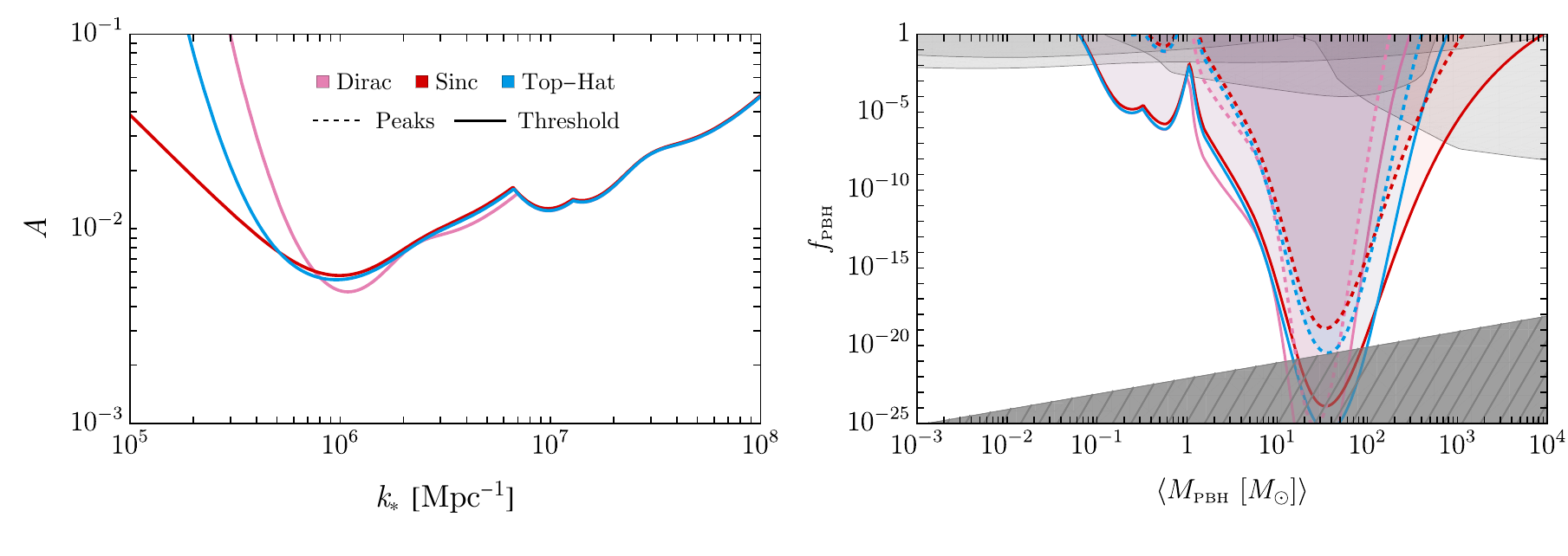}
    \caption{\justifying \textit{Left panel:}
Constraints from NANOGrav15~\cite{NANOGrav:2023gor} on the amplitude of a log-normal curvature power spectrum \eqref{eq:PSlog} with $\Delta=0.5$ assuming Gaussian fluctuations and changing the window function. The horizontal lines correspond to $f_{\rm PBH}=1$ using threshold statistics (solid) and theory of peaks (dashed), respectively, assuming NGs arising solely from the non-linear corrections.
\textit{Right panel:} Inferred constraints on PBH abundance.}
\label{fig:GWfPBHMF}
\end{figure*}

\section{Constraints on PBHs}\label{sec:Anal}
\subsection{PTA observations}

Consider the constraints on the SIGW signal from the current PTA observations. Using from the NANOGrav 15 year dataset~\cite{NANOGrav:2023gor,the_nanograv_collaboration_2023_10344086}, 
we compute the observational upper bound at 95\%  confidence level on $h^2\Omega^{\rm bound}_{\rm GW}$ in each of the first 14 frequency bins. The frequency of bin "$i$" is given by $f_i = i/T$, where $T =  16.03\,{\rm yr} = (1.98 \,{\rm nHz})^{-1}$ is the time of observation. We then compare the observation with theoretical prediction $h^2\Omega_{\rm GW}$ in Eq.~\eqref{eq:total_SIGW} in each frequency bin. The constraint on $h^2\Omega_{\rm GW}$ in any given bin implies an upper bound on $A$ for a given model of NGs and a fixed shape of the curvature power spectrum. To combine the constraints of individual bins, we consider the strongest constraint on $A$. The upper bound on $A$ is then mapped to an upper bound on $f_{\rm PBH}$. More specifically, for every fixed benchmark model in Table \ref{tab:1}, we can map the pair $(A,k_*)$ uniquely into the pair $(f_{\rm PBH},\langle M_{\rm PBH}\rangle)$.

Due to the finite observational time, the spectral resolution of the detector is limited and can be estimated to be approximately $\Delta f = 1/T = 1.98 \,{\rm nHz}$. This will limit the PTA sensitivity to sharp features in the SIGW spectrum. To account for this, the contribution to the $i$-th bin can be estimated as
\be
    \Omega_{{\rm GW},i} = \int \frac{\td f}{f} \Omega_{\rm GW}^{\rm th}(f) W(f-f_i) \,
\ee
where $W$ denotes a window function. To understand the effect of limited spectral resolution, we will consider three different idealized cases
\begin{align}
    W_{\rm D}(f) &\propto \delta \left(f\right)\,,  \\
    W_{\rm S} (f) &\propto {\rm sinc}^2\left( \pi \frac{f}{\Delta f}\right) \,, \\
    W_{\rm TH} (f) &\propto \theta\left(f+\frac{\Delta f}{2}\right) \theta\left(\frac{\Delta f}{2} - f \right) \,,
\end{align}
which correspond to an infinitely accurate frequency resolution, a sharp binning of frequencies and a time-domain top-hat filter on $h_{\rm c}(t)$ with duration $T$, respectively. The window functions are normalized so that $\int \td \ln f \, W(f-f_i) = 1$. These window functions represent different limiting cases and can thus capture systematics related to the spectral resolution not otherwise accessible in our simplified analysis. As seen from Fig.~\ref{fig:GWfPBHMF}, the effect is expectedly larger at large scales and for heavy PBHs as these are mostly related to the signal at small frequencies. For instance, at $k\simeq10^{6}$ $ {\rm{Mpc}}^{-1}$, $W_{\rm D}(f)$ gives a mildly stronger constraint on $A$ when compared to the other choices, as it can resolve sharp peaks in the SIGW spectrum. This is reflected in the constraints on the PBH abundance, as shown on the right panel of Fig.~\ref{fig:GWfPBHMF}. This behaviour is reverted at even smaller $k_*$, and $W_{\rm D}(f)$ is less constraining than the other choices. For the remainder of the analysis, we will consider the Dirac window function $W_{\rm D}(f)$.

\begin{figure*}
    \centering
    \includegraphics[width=0.99\textwidth]{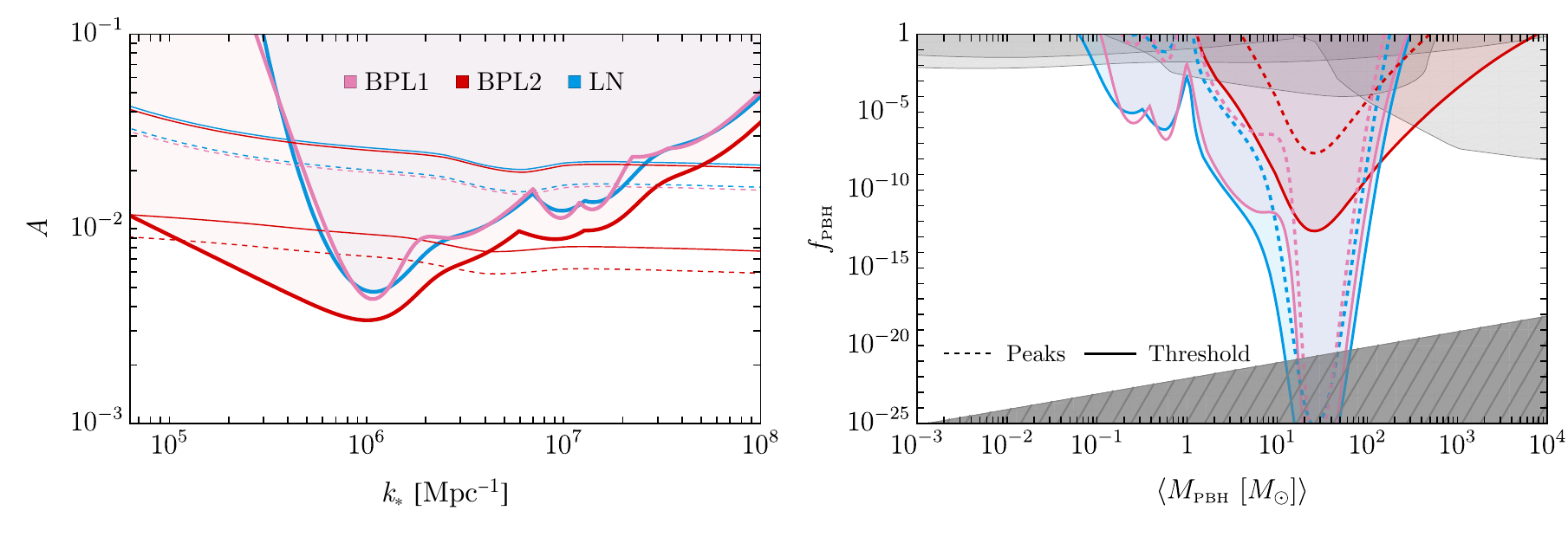}
    \caption{\justifying \textit{Left panel:}
Constraints from NANOGrav15~\cite{NANOGrav:2023gor} on the amplitude of a log-normal curvature power spectrum, Eq.~\eqref{eq:PSlog}, with $\Delta=0.5$ (blue), and on a broken power-law curvature power spectrum, Eq.~\eqref{eq:PPL}, with $\alpha = 4$, $\gamma = 1$ for $\beta = 3$ (pink) and $\beta = 0.5$ (red),  assuming Gaussian fluctuations. The horizontal lines correspond to $f_{\rm PBH}=1$ using threshold statistics (solid) and theory of peaks (dashed), respectively, given the non-linearity-only scenario.
\textit{Right panel:} The implied NANOGrav15 constraints on PBH abundance.}
    \label{fig:GWfPBH}
\end{figure*}

\begin{figure*}
    \centering
    \includegraphics[width=0.99\textwidth]{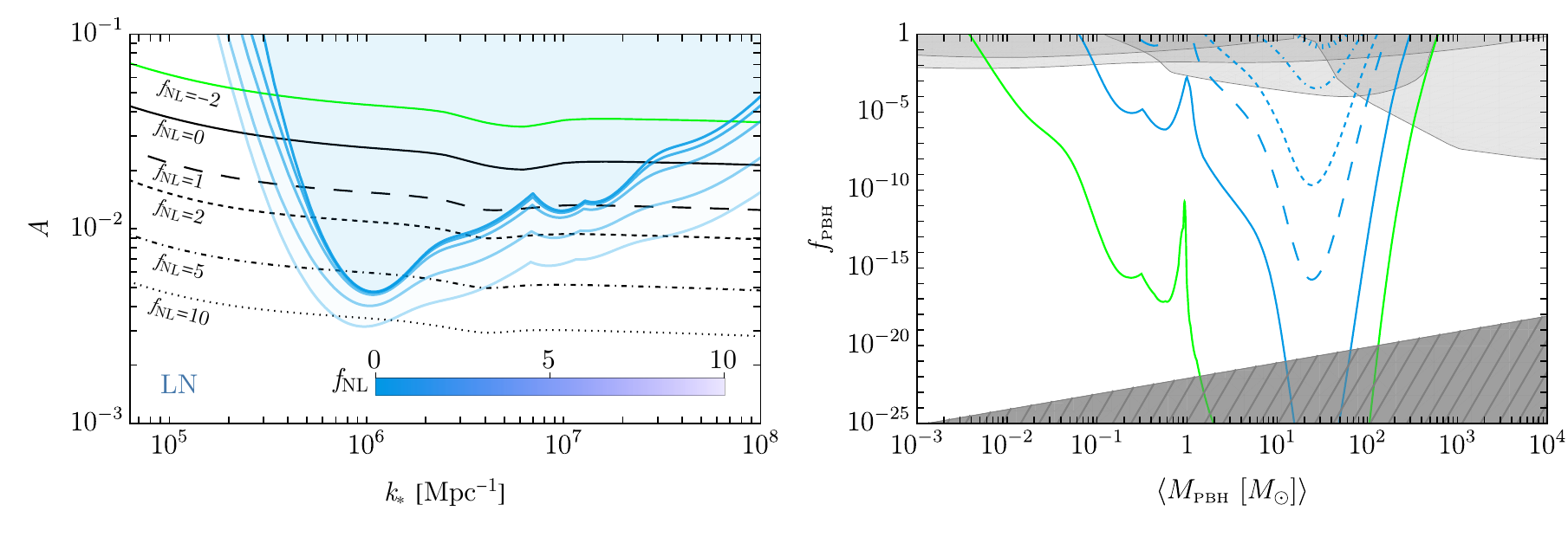}
    \vspace{5mm}
    \includegraphics[width=0.99\textwidth]{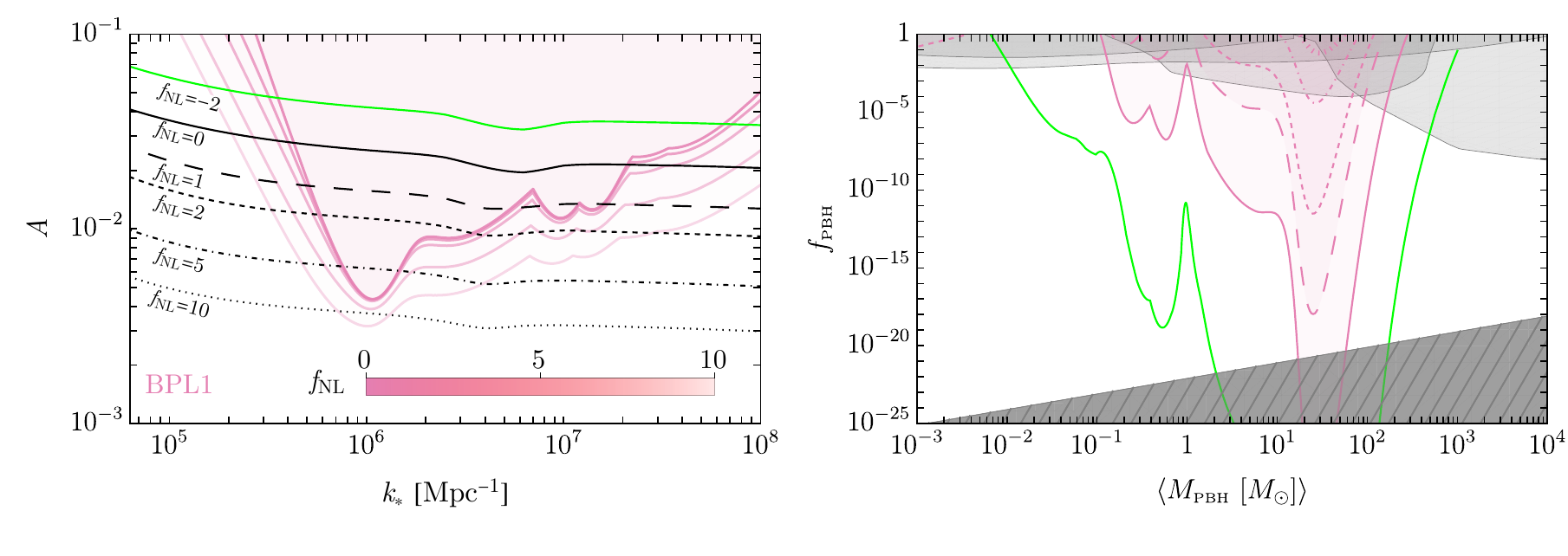}
    \vspace{1mm}
    \includegraphics[width=0.99\textwidth]{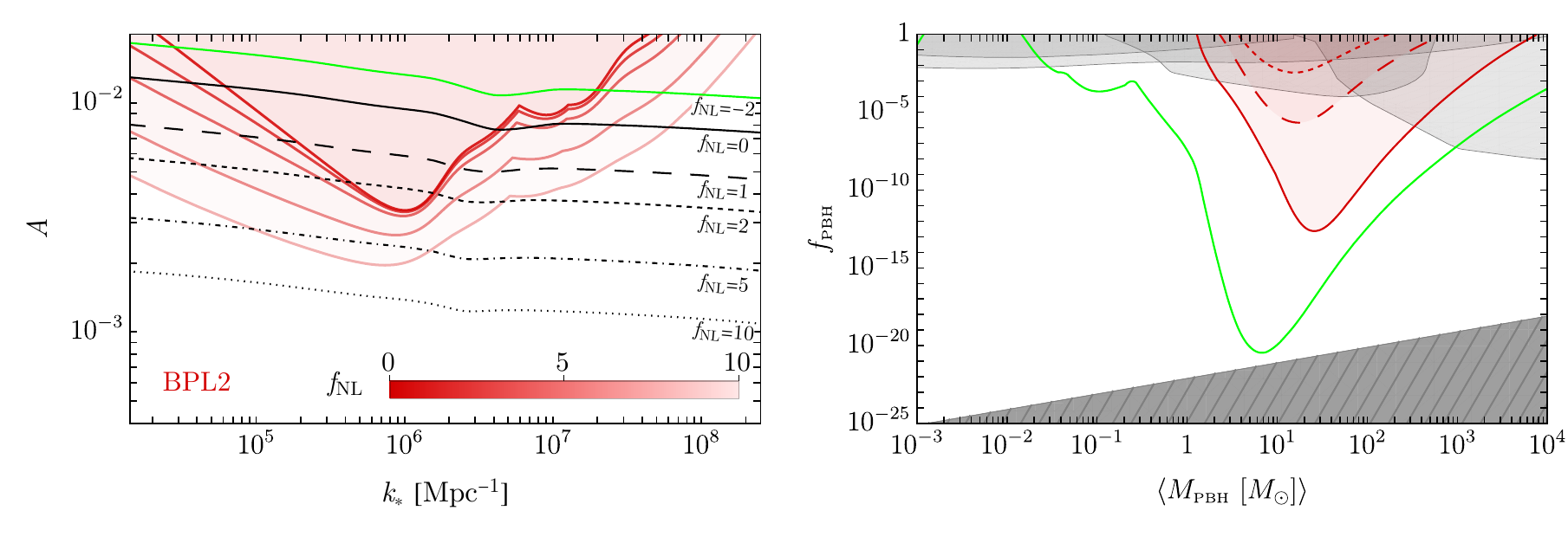}
    \caption{\justifying 
\textit{Left panels:} Constraints from NANOGrav15~\cite{NANOGrav:2023gor} on the amplitude of a log-normal curvature power spectrum \eqref{eq:PSlog} with $\Delta=0.5$ (top panels) and on a broken power-law curvature power spectrum \eqref{eq:PPL} with $\alpha = 4$, $\gamma = 1$ for $\beta = 3$ (middle panels) and $\beta = 0.5$ (bottom panels), assuming quadratic primordial NGs only with $|f_{\rm NL}|\in [0,10]$. The nearly horizontal lines correspond to $f_{\rm PBH}=1$ using threshold statistics with several $f_{NL}$ values.
\textit{Right panels:} Inferred constraints on the PBH abundance using threshold statistics.
}
\label{fig:GWfPBHNGs}
\end{figure*}

Another important theoretical uncertainty when estimating constraints on the PBH abundance arises due to the choice of the prescription when computing the PBH abundance. The right panels of Figs.~\ref{fig:GWfPBHMF} and~\ref{fig:GWfPBH} show the constraints on the PBH abundance arising when the abundance is computed using threshold statistics (solid lines) and peaks theory (dashed lines). In both figures, primordial NGs are neglected ($f_{\rm NL} = 0$) so the NGs arise only from the non-linear relation between density and curvature perturbations. As one can see, almost the entire parameter space accessible by LVK via GWs from PBH binary mergers~\cite{Hutsi:2020sol, Romero-Rodriguez:2021aws, Franciolini:2022tfm, Andres-Carcasona:2024wqk} is excluded when threshold statistics is assumed. On the other hand, with the peaks theory, the constraints are relaxed of few orders of magnitude. Despite this significant theoretical uncertainty, PTA observations can constrain the PBH abundance in both cases.

The dependence of the constraints on the shape of the power spectrum is illustrated in Fig.~\ref{fig:GWfPBH}. While relatively narrow spectra, such as the LN and the BPL1 cases, produce similar results, broader spectra, represented by the BPL2 case, are constrained also at larger scales due to the broadness of the spectrum. We remark that the constraints at large scales $k_* \lesssim 5 \times 10^6 {\rm Mpc}^{-1}$ arise mostly from the first bin by NANOGrav15. This is also the region in which the SIGW are most strongly constrained. In a similar vein, the constraints on the PBH abundance in the threshold statistics case are quite similar for all spectra except at large masses $\langle M_{\rm PBH} \rangle$. In particular, the strongest constraints touch the dashed region in Fig.~\ref{fig:GWfPBH} which corresponds to having less than a single PBH within the current Hubble volume. This line is given by $f_{\rm PBH} 4\pi \Omega_{\rm DM} M_{\rm pl}^2/H_0 < \langle M_{\rm PBH}\rangle$.

Fig.~\ref{fig:GWfPBHNGs} shows the effect of primordial NGs. Firstly, we underline that the NG corrections to SIGW depend on $f_{\rm NL}^2$ and are thus indifferent to the sign of $f_{\rm NL}$. On the other hand, this latter strongly impacts  $f_{\rm PBH}$. Indeed, negative primordial NGs suppress the tail of the PDF with respect to the Gaussian case, that is, they reduce the probability of large fluctuations that cross the threshold. So, a larger amplitude would be needed to obtain the same PBH abundance. Positive primordial NGs, on the other hand, have the opposite effect as they make large fluctuations more likely. 

As shown on the left panels of Fig.~\ref{fig:GWfPBHNGs}, for small $f_{\rm NL}$ we observe that the corrections to the SIGW affect only amplitudes at the tails, both at smaller and higher scales. Larger NGs ($f_{\rm NL} \geq 5$), on the other hand, shift the entire SIGW constraint curves. The effect on the $f_{\rm PBH}$ constraints, shown in the right panels of Fig.~\ref{fig:GWfPBHNGs}, can be qualitatively understood by comparing the shift of $f_{\rm PBH} = 1$ curves when compared to the SIGW constraint: Since the effect on $f_{\rm PBH} = 1$ is larger than the strengthening of the SIGW constraints, positive primordial NGs can completely remove constraints on the PBH abundance. We find that for $f_{\rm NL} = 5$, the PTA constraints on $f_{\rm PBH}$ are eliminated in the wide BPL2 case and become weaker than the constraints from other observables in the narrow LN and BPL1 cases.

The opposite is observed with negative NGs -- since they strengthen the constraints on SIGW in a similar way to positive NGs, but require a higher amplitude $A$ to achieve the same PBH abundance, the constraints on $f_{\rm PBH}$ become more stringent. This is indicated by the green line in Fig.~\ref{fig:GWfPBHNGs}. In particular, even a mild negative NG ($f_{\rm NL} = -2$) could exclude the existence of any PBHs in our present Hubble volume in the mass range $2-100 M_{\odot}$, given a narrow spectrum as in the LN and BPL1 cases.

However, one might mistakenly assume that by increasing the negative value of the coefficient $f_{\rm NL}$, the required power spectral amplitude for the same PBH abundance would always rise, tightening the constraints indefinitely. However, this is not the case, as the required amplitude reaches a maximum at $f_{\rm NL} \simeq -2$ and then decreases, eventually approaching the amplitude of the Gaussian case for very large negative values of $f_{\rm NL}$~\cite{Franciolini:2023pbf}. Consequently, the constraints for the case of $f_{\rm NL} = -2$ can be considered the most stringent.

Nevertheless, this claim should be taken \textit{cum grano salis}. Indeed, it is important to note that the prescription outlined in Ref.~\cite{Musco:2020jjb} to compute the threshold for PBH collapse only accounts for NGs arising from the non-linear relation between the density contrast and the curvature perturbations. In principle, also primordial NGs beyond the quadratic approximation should be taken into account when computing the threshold value. Following Refs.~\cite{Kehagias:2019eil, Escriva:2022pnz}, it appears that their effect on the threshold is small when the NGs are positive or mildly negative ($f_{\rm NL} \gtrsim -2$). In such cases, the threshold receives corrections of at most a few percent. For stronger negative NGs, the uncertainties on the exact value of the threshold can be sizeable, and could thus potentially modify the constraints introduced above\footnote{One alternative approach is to use the averaged value over a sphere of radius equal to the location of the maximum of the compaction function, $\Bar{C}_{\rm th} = 2/5$, since this value is fully non-perturbative and independent of NGs~\cite{Escriva:2019phb, Kehagias:2024kgk}. As discussed in Ref.~\cite{Ianniccari:2024bkh}, this requires determining the statistics of the curvature perturbation and computing the connected cumulants, which is a complex task and is left for future work.}. 

Although evaluating the NG corrections to the formation threshold is beyond the scope of this work, we can estimate the impact on the PBH abundance due to a few percent correction on $\mathcal{C}_{\rm th}$. In detail, we re-computed the constraints on $f_{\rm PBH}$ considering a 5\% change in $\mathcal{C}_{\rm th}$ for the three benchmark models in Table~\ref{tab:1}. In all the cases, the constraints on the abundance were affected by at most two orders of magnitude. Therefore, in the light of Fig.~\ref{fig:GWfPBHNGs}, we expect that perturbative corrections on $\mathcal{C}_{\rm th}$ would not induce significant changes to the PTA constraints on $f_{\rm PBH}$ and thus our main conclusions would not be strongly affected.

\begin{figure}[t]
    \centering
    \includegraphics[width=0.85\textwidth]{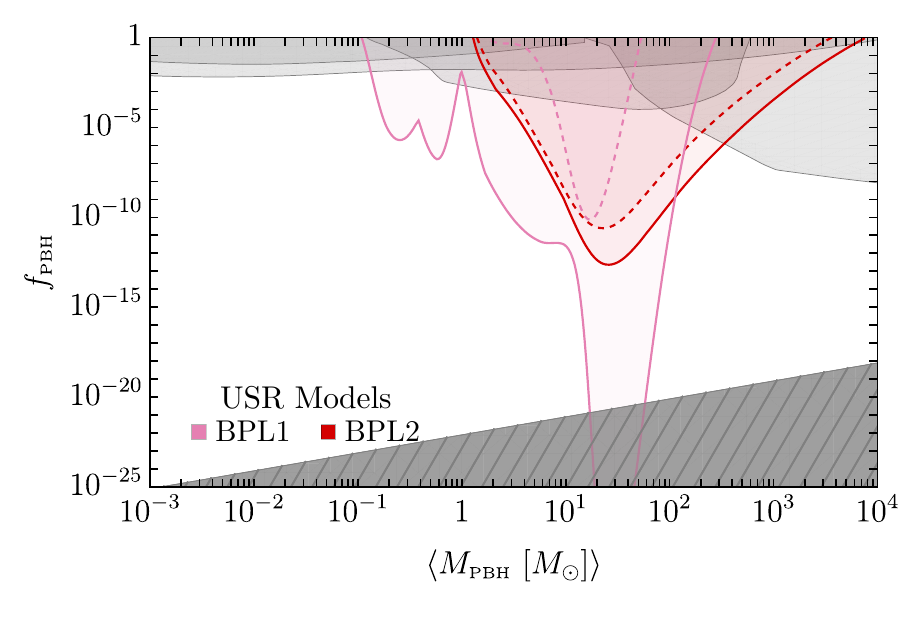}
    \caption{\justifying 
The NANOGrav15 constraints on the PBH abundance using threshold statistics assuming the two cases for the broken power law reported in Tab.~\ref{tab:1} assuming the Gaussian approximation (solid) and with the full NG relation in Eq.~(\ref{eq:zeta_IP}).
}\label{fig:USR}
\end{figure}

For completeness, we analyze a specific but common scenario for PBH formation. 
In \emph{Ultra slow roll} (USR) models of single-field inflation, the peak in $\mathcal{P}_{\zeta}$ arises from a brief phase of ultra-slow-roll which is typically followed by slow-roll or constant-roll inflation dual to it~\cite{Atal:2018neu, Biagetti:2018pjj, Karam:2022nym}. In this case, the NGs can be related to the large $k$ spectral slope generated during the last inflationary phase~\cite{Atal:2019cdz, Tomberg:2023kli},
\be\label{eq:zeta_IP}
    \zeta = -\frac{2}{\beta}\log\left(1-\frac{\beta}{2}\zeta_{\rm G}\right).
\ee
In Fig.~\ref{fig:USR}, we show the abundance of PBHs using the full NG relation Eq.~\eqref{eq:zeta_IP} for the two broken power law cases reported in Tab.~\ref{tab:1}.  The PBH abundance is computed using threshold statistics and compared with the Gaussian approximation. As seen from Fig.~\ref{fig:USR}, NGs can significantly affect the constraints on $f_{\rm PBH}$ when the spectrum is narrow (in the BPL1 case). For wider spectra (BPL2 case), the NG-induced suppression is milder and thus the final constraint is stronger. To make contact with the $f_{\rm NL}$ description of NGs, expanding Eq.~\eqref{eq:zeta_IP} yields $f_{\rm NL} = 5 \beta/12 $, so that $\beta = 3$ and $\beta=0.5$ correspond to $f_{\rm NL} = 5/4$ and $f_{\rm NL}= 5/24$, thus the NGs are relatively small and the NG corrections to SIGW have a subdominant effect. Consequently, for USR models, PTAs are more effective in constraining PBHs from wide curvature power spectra due to the weaker NG-induced suppression.

The other constraints reported here are GW O3~\cite{Andres-Carcasona:2024wqk}, EROS~\cite{EROS-2:2006ryy}, OGLE~\cite{Mroz:2024mse,Mroz:2024wag}, Seg1~\cite{Koushiappas:2017chw}, Planck~\cite{Serpico:2020ehh,Agius:2024ecw, Facchinetti:2022kbg}, Eri II~\cite{Brandt:2016aco}, WB~\cite{Monroy-Rodriguez:2014ula}, Ly$-\alpha$~\cite{Murgia:2019duy} and SNe~\cite{Zumalacarregui:2017qqd}. These constraints are shown for monochromatic mass functions.

\begin{figure*}
\centering
\includegraphics[width=0.85\textwidth]{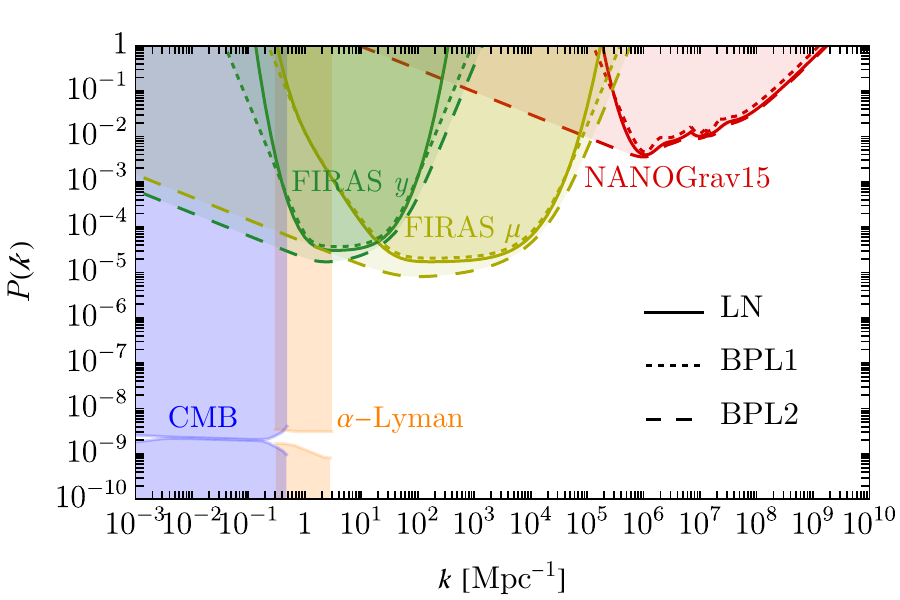}
\caption{\justifying Constraints on the amplitude of the power spectrum, assuming negligible primordial NGs, from the FIRAS experiments, CMB, $\alpha$-Lyman and NANOGrav15 for the same cases reported in Tab.~\ref{tab:1}.}
\label{fig:muconstra}
\end{figure*}

\subsection{$\mu$ distortions and PBH seeds for SMBHs}

At the largest scales, the primordial power spectrum is strongly constrained by the CMB observations~\cite{Planck:2018vyg} and the Lyman-$\alpha$ forest data~\cite{lyman}\footnote{For recent works on how future CMB experiments can probe the range of scales related to PTA see Refs.~\cite{Cyr:2023pgw,Tagliazucchi:2023dai}}.
The CMB observations strongly constrain the curvature power spectrum at scales $10^{-4} \mathrm{Mpc}^{-1} \lesssim k \lesssim 1 \mathrm{Mpc}^{-1}$. 
At redshifts $z \lesssim 10^6$, energy injections into the primordial plasma cause persisting spectral distortions in the CMB. These distortions are divided into chemical potential $\mu$-type distortions created at early times and Compton $y$-type distortions created at $z \lesssim 5 \times 10^4$.
For a given curvature power spectrum $\mathcal{P}_{\zeta}(k)$ the spectral distortions are~\cite{Chluba:2012we,Chluba:2013dna}
\be
    X=\int_{k_{\rm min}}^{\infty} \frac{\mathrm{d} k}{k} \mathcal{P}_{\zeta}(k) W_X(k)
\ee
with $X=\mu, y$, while $k_{\rm min}=1$ Mpc$^{-1}$ and the window functions can be approximated by
\be
    W_\mu(k)=2.2\left[e^{-\frac{(\hat{k} / 1360)^2}{1+(\hat{k} / 260)^{0.6}+\hat{k} / 340}}-e^{-(\hat{k} / 32)^2}\right], 
\ee
\be
W_y(k)=0.4 e^{-(\hat{k} / 32)^2}
\ee
with $\hat{k}=k /\left(1 \mathrm{Mpc}^{-1}\right)$.
The COBE/Firas observations constrain the $\mu$  distortions as $\mu \leq 4.7 \times 10^{-5}$~\cite{Bianchini:2022dqh} and $y\leq 1.5 \times 10^{-5}$ at $95\%$ confidence level~\cite{Fixsen:1996nj}.
The situation is summarised in Fig.~\ref{fig:muconstra}.
Including NGs modifies only the tails of the constraints and not the most constrained region~\cite{Sharma:2024img}.

Consequently, the amplitude of the power spectrum in the range of scales related to the FIRAS experiment is constrained to be at most $A=10^{-5}$.

We focus on the threshold statistics approach and we compute the mass fraction $\beta$ as a function of the parameters $f_{\rm NL}$ and $g_{\rm NL}$, assuming the power series ansatz as in Eq.~\ref{eq:exp}. The power series expansion holds until we are in the perturbative regime, that is, the terms are ordered hierarchically with the higher orders being typically smaller than the lower ones. For a narrow power spectrum, this translates into a constraint for the amplitude of the power spectrum, $(3/5)\left|f_{\mathrm{NL}}\right| A^{1 / 2} \ll 1$, and $(9/25)\left|g_{\mathrm{NL}}\right| A \ll 1$.

Such large positive NGs can lead to the production of a non-negligible population of massive PBHs that can seed SMBHs.

\begin{figure*}
    \centering
\includegraphics[width=0.99\textwidth]{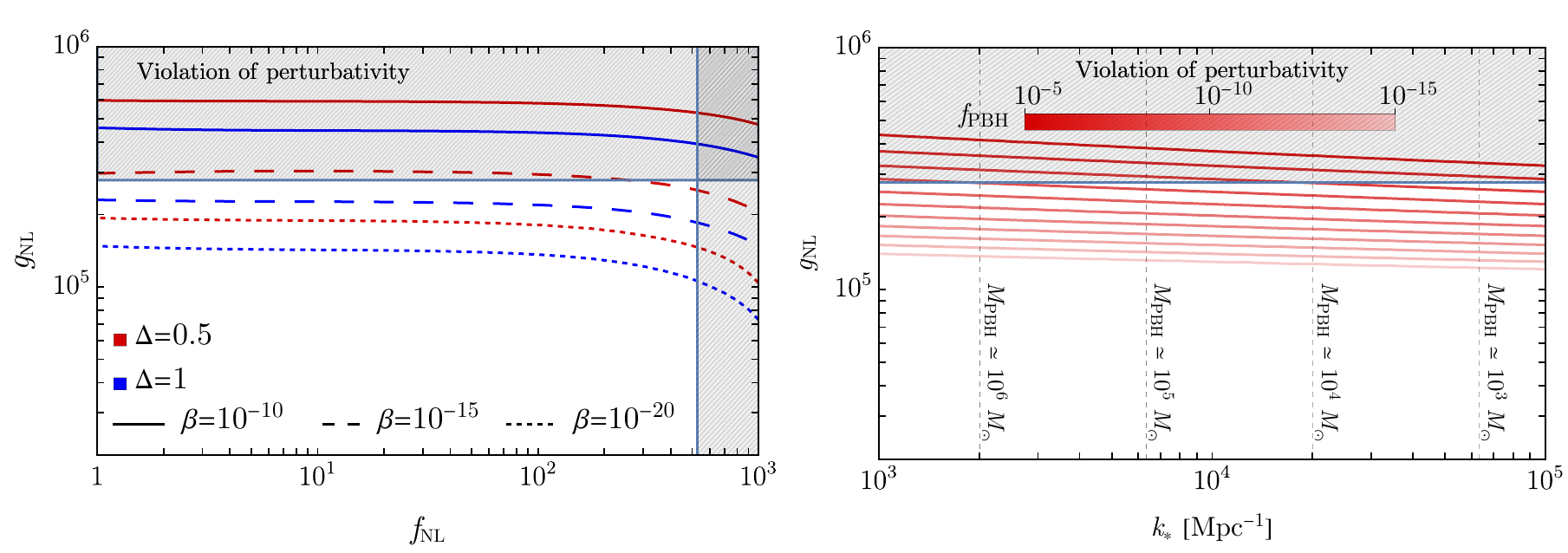}
\caption{\justifying \textit{Left panel:} Computation of the mass fraction $\beta$ changing the NG parameters $f_{\rm NL}$ and $g_{\rm NL}$ with two benchmark cases for the log-normal power spectrum with an amplitude fixed to be $A=10^{-5}$. \textit{Right panel:} Computation of the PBH abundance $f_{\rm PBH}$ changing the scale of the peak $k_*$ and the NG parameter $g_{\rm NL}$. We fix $\Delta=1$, $A=10^{-5}$ and $f_{\rm NL} = 1$.}
\label{fig:fpbhgnl}
\end{figure*}
Indeed, the origin of SMBHs presents a significant challenge in the field of astrophysics. Although it is well known that they occupy the centres of most galaxies~\cite{Kormendy:1995er, Magorrian:1997hw, Richstone:1998ky}, the processes leading to their formation remain unclear.

In the context of PBHs, even a minimal presence of massive PBHs within the mass range $10^3-10^6 \, M_{\odot}$ could potentially seed for SMBHs~\cite{Duechting:2004dk,Kohri:2014lza,Bernal:2017nec}. As a naive computation to estimate the necessary abundance of these seeds, we follow Refs.~\cite{Vaskonen:2020lbd, Serpico:2020ehh}. Considering that SMBHs make up approximately $\mathcal{O}(10^{-4})$ of the stellar mass in their host galaxies~\cite{Reines_2015}, and that stars account for roughly $\mathcal{O}(10^{-2})$ of the total cosmic matter content~\cite{Fukugita:1997bi}, we deduce that the overall SMBH density is about $10^6$ times less than the dark matter density.

The primordial seeds' abundance can then be expressed as
\be\label{eq:SMBH}
    f_{\mathrm{PBH}} \sim \mathcal{O}(10^{-6}) \times \left\langle M_{\mathrm{PBH}}\right\rangle / M_{\rm SMBH}\, ,
\ee
with $\left\langle M_{\mathrm{PBH}}\right\rangle > 10^3 M_{\odot}$.
Nevertheless, the scales related to the formation of SMBHs are strongly constrained by the analysis of the CMB spectral distortion by the FIRAS collaboration~\cite{Fixsen:1996nj,Chluba:2012gq,Chluba:2012we,Chluba:2013dna,Bianchini:2022dqh}, and in the gaussian approximation, the corresponding abundance of PBHs is too small to furnish a primordial origin for Supermassive black holes seeds. 

As shown in the left panel of Fig.~\ref{fig:fpbhgnl}, in agreement with the literature~\cite{Unal:2020mts,Nakama:2017xvq,Hooper:2023nnl,Byrnes:2024vjt}, quadratic primordial NGs are not enough to produce a sizeable amount of PBHs to be the seed of SMBHs. Consequently, differently from Ref.~\cite{Byrnes:2024vjt}, we find that already a sizeable cubic parameter $g_{\rm NL}$ is enough to overcome this hurdle. This discrepancy arises because their analysis is subject to a few simplifications: using the curvature perturbation instead of the density contrast to determine the collapse and neglecting the threshold's dependence on the shape of the curvature power spectrum.

According to Fig.~\ref{fig:fpbhgnl}, assuming a typical SMBH mass of approximately $10^{8}$ $M_{\odot}$, condition~\eqref{eq:SMBH}, without the violation of the perturbative criterium, holds across the range of PBH masses analyzed.

\section{Conclusions and outlook}
\label{sec:Conc}

In this work, we have updated the constraints on the amplitude of the curvature power spectrum, ensuring that the amplitude of the produced SIGWs does not exceed the signal recently detected by the NANOGrav collaboration.

When PBHs are formed via the collapse of sizeable curvature perturbations, the constraints on SIGW will infer an upper bound on the PBH abundance. We derived these bounds in various scenarios and by estimating PBH abundance using threshold statistics as well as the theory of peaks. 

We found that, when using threshold statistics, only a small fraction of dark matter in the form of PBHs is available in the solar mass range regardless of the specific shape of the power spectrum. For narrow curvature power spectra, significant constraints $f_{\rm PBH}\lesssim 10^{-5}$ arise in the mass range $0.1-100 M_{\rm \odot}$, while for broader spectra, such constraints arise for heavier $5-500 M_{\rm \odot}$ PBHs.

These potential constraints can, however, be significantly relaxed in the presence of positive NGs, and are eliminated when $f_{\rm NL} \gtrsim 10$. However, we find that stellar mass PBHs tend to remain constrained when considering NGs present in typical USR models for PBH production. In these cases, wider spectra tend to infer strong constraints over a wider PBH mass range as they are associated with weaker NGs.

When the PBH abundance is estimated using the theory of peaks, however, constraints on $f_{\rm PBH}$ are relaxed of few orders of magnitude.

Finally, we discussed how large primordial NGs ($g_{\rm NL} \simeq 10^{5}$) permit the production of PBH seeds for SMBHs without violating perturbativity. Despite this possibility, the literature currently lacks explicit models capable of producing sufficiently large NGs. Indeed, in common scenarios, such as the curvaton and USR models, the general impact of the NGs is curtailed when one goes beyond the perturbative approach~\cite{Ferrante:2022mui, Ferrante:2023bgz}. 

To improve future estimates, it is crucial to refine the computation of the PBH abundance, which is still subject to a series of theoretical uncertainties.

\acknowledgments
\noindent
We thank G. Franciolini and X. Pritchard for useful discussions and A. Ianniccari, D. Perrone and V. Vaskonen for comments on the draft.
A.J.I. acknowledges the financial support provided under the “Progetti per Avvio alla Ricerca Tipo 1”, protocol number AR1231886850F568.
H.V. is supported by the Estonian Research Council grants PSG869 and RVTT7 and the Center of Excellence program TK202.

\clearpage

\bibliographystyle{JHEP}
\bibliography{refs}

\end{document}